\documentclass[aps, prl, 10pt, twocolumn, showpacs, superscriptaddress]{revtex4-2}

\usepackage[utf8]{inputenc}
\usepackage{amsmath,amssymb,amsfonts,amsthm,amsopn,amscd,bm}
\usepackage{dsfont}
\usepackage{multirow}
\usepackage{graphicx}
\usepackage{booktabs}
\renewcommand{\arraystretch}{1.2}
\usepackage[shortlabels]{enumitem}
\usepackage{xcolor}
\usepackage{enumitem}
\usepackage[normalem]{ulem}
\usepackage{kbordermatrix}
\usepackage{ifthen}
\usepackage{tikz}
\usetikzlibrary{cd}
\usetikzlibrary{positioning,arrows, calc}
\usetikzlibrary{decorations.pathmorphing}
\usetikzlibrary{decorations.markings}

\colorlet{gold}{yellow!80!red!80!black}

\usepackage[breaklinks]{hyperref}
\hypersetup{colorlinks,
    linkcolor={red!40!black},
    citecolor={blue!50!black},
    urlcolor={blue!20!black}
}
\usepackage{orcidlink}

\tikzset{
 overlap/.style args = {#1/#2}{minimum height=#1,
               minimum width=#2,
               rounded corners=2.2mm, fill=black!50, opacity=0.6, sloped},
               party/.style = {draw, circle, fill=black, minimum width=1mm},
obi/.style={black!50, decorate, decoration={
            snake,
            segment length=2mm,
            amplitude=0.2mm}
            }
}

\tikzset{shifted path/.style args={from #1 to #2 by #3}{insert path={
let \p1=($(#1.east)-(#1.center)$),
\p2=($(#2.east)-(#2.center)$),\p3=($(#1.center)-(#2.center)$),
\n1={veclen(\x1,\y1)},\n2={veclen(\x2,\y2)},\n3={atan2(\y3,\x3)} in
(#1.{\n3+180+asin(#3/\n1)}) to (#2.{\n3-asin(#3/\n2)})
}}}

\DeclareMathOperator{\tr}{tr}
\DeclareMathOperator{\one}{\mathds{1}}

\newcommand{\ket}[1]{\mathinner{|#1\rangle}}

\newcommand{\ot}[0]{\otimes}
\renewcommand{\a}{\alpha}
\renewcommand{\b}{\beta}
\newcommand{\expect}[1]{\langle #1 \rangle}

\newcommand{\N}{\mathds{N}}

\newcommand{\C}{\mathds{C}}

\newcommand{\nn}{\nonumber}

\newcommand{\id}{\operatorname{id}}

\newcommand{\prom}[1]{\langle #1 \rangle}

\newcommand{\LL}{\mathcal{L}}

\newcommand{\comp}{\beta}

\newcommand{\Ibound}{\nu_\ell}
\newcommand{\state}[1]{\langle #1\rangle_\varrho}
\newcommand{\statesym}[1]{\state{#1}}
\renewcommand{\statesym}[1]{\expect{#1}}
\renewcommand{\ss}[1]{\statesym{#1}}

\newlength{\minuslength}
\begin{document}

\title{Uncertainty relations from state polynomial optimization}

\author{Mois\'es Bermejo Mor\'an${}^{\orcidlink{0000-0003-1441-0468}}$}
\affiliation{
Faculty of Physics, Astronomy and Applied Computer Science, Institute of Theoretical Physics, Jagiellonian University,
30-348 Krak\'{o}w, 
Poland}
\email{moises.moran@uj.edu.pl}

\author{Felix Huber${}^{\orcidlink{0000-0002-3856-4018}}$}
\affiliation{
Faculty of Physics, Astronomy and Applied Computer Science, Institute of Theoretical Physics, Jagiellonian University,
30-348 Krak\'{o}w, 
Poland}
\affiliation{
Bordeaux Computer Science Laboratory (LaBRI),
University of Bordeaux,
351 cours de la Liberation,
33405 Talence, France}

\date{\today}

\begin{abstract}
Uncertainty relations are a fundamental feature of quantum mechanics.
How can these relations be found systematically?
Here we develop a semidefinite programming hierarchy
for additive uncertainty relations in the variances of non-commuting observables.
Our hierarchy is built on the state polynomial optimization framework,
also known as scalar extension.
The hierarchy is complete,
in the sense that it converges to tight uncertainty relations.
We improve upon upper bounds for all 1292 additive uncertainty relations
on up to nine operators
for which a tight bound is not known.
The bounds are dimension-free and
depend entirely on the algebraic relations among the operators.
The techniques apply to a range of scenarios,
including Pauli, Heisenberg-Weyl,
and fermionic operators, and generalize to higher order moments
and multiplicative uncertainty relations.
\end{abstract}

\maketitle
\section{Introduction}

\label{sec:intro}
Quantum mechanical particles exhibit
a fundamental uncertainty relation between conjugate observables.
Similar to probability distributions that are related by a Fourier transform,
conjugate observables do not allow for simultaneous measurements with vanishing variances.
This feature both enables as well as hinders certain applications of
quantum technologies~\cite{fuchs1996quantum, guhne2004characterizing, cavalcanti2007uncertainty, renes2009conjectured, gottesman2010introduction, bennett2014quantum}.
While commuting measurements can be jointly diagonalized, anti-commuting observables are incompatible:
This is taken advantage of in quantum error correction,
where errors that anti-commute with some stabilizer elements are
detected by the change they induce in the respective error syndromes~\cite{gottesman2010introduction}.
Also the security of quantum key distribution relies crucially
on the fact that a basis change leads to an unavoidable uncertainty in measurement outcomes~\cite{bennett2014quantum}.
In particular, additive uncertainty relations find applications in spectroscopy and atomic clocks~\cite{sorensen2001}, and quantum metrology~\cite{toth2022}.
Considerable effort has been made to derive state-independent bounds~\cite{dammeier2015, szymanski2019}.

To introduce our setting,
consider a two-dimensional spin system with $\sigma_x, \sigma_y, \sigma_z$ the Pauli matrices
and denote $\langle \sigma_i \rangle_\varrho :=
\tr(\varrho \, \sigma_i )$ the expectation value of $\sigma_i$ on state $\varrho$.
The set of a qubit density matrices is described by the Bloch ball,
satisfying
\begin{equation}
\langle \sigma_x \rangle_\varrho^2 + \langle \sigma_y \rangle_\varrho^2 + \langle \sigma_z \rangle_\varrho^2 \leq 1\,.
\label{eq:xyz}
\end{equation}
A straightforward consequence of this constraint is an additive {uncertainty} relation of the form
\begin{equation}\label{eq:uncerainty_qubit}
  \Delta^2 \sigma_x + \Delta^2 \sigma_y + \Delta^2 \sigma_z \geq 2\,,
\end{equation}
where
$\Delta^2 \sigma_i := \langle \sigma_i^2\rangle_\varrho - \langle \sigma_i\rangle_\varrho^2$ and we used that
$\sigma_i^2 = \one$.
More generally, for a set of anti-commuting operators the relation $\sum_{i=1}^n \langle \sigma_i \rangle^2 \leq 1$ holds. This follows from the decomposition 
$
1 - \sum \langle {\sigma_i} \rangle_\varrho^2 = \langle
( 1-\sum \sigma_i \langle \sigma_i \rangle_\varrho )^2\rangle \geq 0 \, $ as sum of squares.
A natural question is then what happens when some of the variables commute, while some others anti-commute.

To answer this question we investigate the following quantity:
Given a family of hermitian and unitary operators $\{A_i\}_{i=1}^n$
satisfying commutation relations of the form
$A_i A_j = \pm A_j A_i$,
define the quantity
\begin{equation}\label{eq:C}
 \comp =\sup_\varrho \sum_{i=1}^n \prom{A_i}_\varrho^2\,.
\end{equation}
Here the maximization is over all states $\varrho$ and bounded operators $A_i$ on Hilbert spaces
that support $\varrho$ and the commutation relations between the $A_i$.
Bounds on $\beta$ have 
several applications in quantum information,
including the characterization of entanglement~\cite{de2022uncertainty}, nonlocality~\cite{kurzynski2011correlation},
measurement compatibility~\cite{loulidi2022}, and to estimate ground state energies~\cite{PRXQuantum.5.020318}.
Alternatively, Eq.~\eqref{eq:C} can
in the spirit of Eq.~\eqref{eq:uncerainty_qubit}
tight additive uncertainty relation in the variances
$\Delta^2 A_i := \langle A_i^2\rangle_\varrho - \langle A_i \rangle_\varrho^2$,
\begin{equation}\label{eq:uncert}
 \sum_{i=1}^n \Delta^2 A_i \geq n-\comp\,.
\end{equation}

In a recent work~\cite{de2022uncertainty},
Gois et al.~provided the upper bound
$\beta \leq \vartheta(G)$.
Here $\vartheta(G)$ is the Lov\'asz theta number of the observables anti-commutativity graph.
This graph encodes the observables' relations, with two vertices joined by an edge,
if the corresponding observables anti-commute; i.e. $i \sim j$ if $A_i A_j = - A_j A_i$
and $i \not \sim j$ else.
The related bound by Hastings and O'Donnell~\cite{hastings2022optimizing},
$\langle \sum_{i=1}^n a_i A_i \rangle_\varrho^2 \leq \vartheta(G)$ for all $||a|| \leq 1$,
has been discovered in the context of optimizing fermionic Hamiltonians.
In turn, $\beta$ is lower bounded by the independence number $\a$,
that is the maximal number of disconnected vertices of $G$.
The appearance of the Lov\'{a}sz theta number in this context is intriguing,
as it has already been linked to nonlocality, contextuality,
and quantum zero-error communication~\cite{Cabello_2014, acin2015combinatorial, Acin_2017, duan2013zero}.

The aim of this work is to provide a semidefinite programming hierarchy
that upper bounds Eq.~\eqref{eq:C} and improves upon the Lov\'asz bound.
The first level of our hierarchy coincides with the Lov\'asz theta number,
and thus with the result of Gois et al.
Furthermore, if at any level of the hierarchy the upper bound coincides with $\a$
or the rank loop condition is met,
then the corresponding uncertainty relation in Eq.~\eqref{eq:uncert} is tight.
Numerical tests show that already the second level of our hierarchy improve
upon the previous best upper bounds for all 1292 non isomorphic graphs with up to nine vertices for which $\beta$ is unknown \cite{private_communication}.
In particular,
Table~\ref{tab:graphs-reduced-t3} shows that
the second and third levels often close the gap left by the Lov\'asz theta number to a tight bound.
We also introduce a second semidefinite programming hierarchy [Eq.~\eqref{eq:statepop-relax}] that is complete
and thus converges to $\beta$.
With this we partially answer an open question by Xu et al.~on
how to obtain efficient upper bounds on $\comp$~\cite{PRXQuantum.5.020318}.

\begin{table}[tbp]
\begin{center}
\def \scale {0.69}
\def \zoom {0.25}
\def \nth {5}
\def \an {360/\nth}
\def \bline {0.1cm}
\def \xrot {198}
\def \htabsep {15}
\def \acolor {black}
\begin{tabular}{c@{\hskip 10pt}@{\hskip 10pt}c@{\hskip \htabsep pt}c@{\hskip \htabsep pt}c@{\hskip \htabsep pt}c}
 & \# 2 & \# 3 & \# 4 & \# 5
 \\
  &
 \begin{tikzpicture}[scale=\scale, baseline=\bline, rotate=\xrot]
 \coordinate (A0) at (0,0);
 \node[party, scale=\zoom] (A0) at (A0) {};
    \foreach \i in {1,2,...,\nth} {
\coordinate (A\i) at ({-cos(\i*\an)},{sin(\i*\an)}) {};
\node[party, scale=\zoom] (A\i) at (A\i) {};
}
\draw (A1) -- (A2) -- (A3) -- (A4) -- (A5) -- (A1);
\node[party, scale=\zoom, \acolor] (A0) at (A0) {};
\node[party, scale=\zoom, \acolor] (A1) at (A1) {};
\node[party, scale=\zoom, \acolor] (A3) at (A3) {};
 \end{tikzpicture}
 &
\begin{tikzpicture}[scale=\scale, baseline=\bline, rotate=\xrot]
 \coordinate (A0) at (0,0);
 \node[party, scale=\zoom] (A0) at (A0) {};
    \foreach \i in {1,2,...,\nth} {
\coordinate (A\i) at ({-cos(\i*\an)},{sin(\i*\an)}) {};
\node[party, scale=\zoom] (A\i) at (A\i) {};
}
\draw (A1) -- (A2) -- (A3) -- (A4) -- (A5) -- (A1);
\draw (A3) -- (A0);
\draw (A4) -- (A0);
\draw (A5) -- (A0);
\node[party, scale=\zoom, \acolor] (A3) at (A3) {};
\node[party, scale=\zoom, \acolor] (A1) at (A1) {};
 \end{tikzpicture}
 &
\begin{tikzpicture}[scale=\scale, baseline=\bline, rotate=\xrot]
 \coordinate (A0) at (0,0);
 \node[party, scale=\zoom] (A0) at (A0) {};
    \foreach \i in {1,2,...,\nth} {
\coordinate (A\i) at ({-cos(\i*\an)},{sin(\i*\an)}) {};
\node[party, scale=\zoom] (A\i) at (A\i) {};
}
\draw (A1) -- (A2) -- (A3) -- (A4) -- (A5) -- (A1);
\draw (A2) -- (A0) -- (A5);
\draw (A1) -- (A0) -- (A3);
\node[party, scale=\zoom, \acolor] (A1) at (A1) {};
\node[party, scale=\zoom, \acolor] (A3) at (A3) {};
 \end{tikzpicture}
 &
\begin{tikzpicture}[scale=\scale, baseline=\bline, rotate=\xrot]
 \coordinate (A0) at (0,0);
 \node[party, scale=\zoom] (A0) at (A0) {};
    \foreach \i in {1,2,...,\nth} {
\coordinate (A\i) at ({-cos(\i*\an)},{sin(\i*\an)}) {};
\node[party, scale=\zoom] (A\i) at (A\i) {};
}
\draw (A1) -- (A2) -- (A3) -- (A4) -- (A5) -- (A1);
\draw (A2) -- (A0) -- (A4);
\draw (A1) -- (A0) -- (A3);
\draw (A0) -- (A5);
\node[party, scale=\zoom, \acolor] (A1) at (A1) {};
\node[party, scale=\zoom, \acolor] (A3) at (A3) {};
 \end{tikzpicture}
 \\
 & \\
 $\vartheta_1$ & 3.2361 & 2.2361 & 2.2361 & 2.2361 \\
 $\vartheta_2$ & 3.0000 & 2.0000 & 2.0000 & 2.0000 \\
 $\alpha$ & 3 & 2 & 2 & 2 
\end{tabular}
\end{center}
\caption{{\bf Tight uncertainty relations.}
The commutation relations of operators are encoded in a graph:
two vertices are connected if the corresponding operators anti-commute
and disconnected otherwise.
Our hierarchy [Eq.~\eqref{eq:lovaszk}]
upper bounds $\beta= \sup_\varrho \sum_{i=1}^n \state{A_i}^2$ as $\b \leq \dots \leq \vartheta_2 \leq \vartheta_1$,
from which one obtains the additive uncertainty relation
$\sum_{i=1}^n \Delta^2 A_i \geq n-\beta$.
The independence number $\alpha$ is the size of the largest set of disconnected vertices
and lower bounds~$\beta$,
while the Lov\'{a}sz number $\vartheta$ [Eq.~\eqref{eq:lovasz}] provides an upper bound and
coincides with the first level of our hierarchy.
For the graphs shown above, the second level gives a tight uncertainty relation as $\alpha = \beta = \vartheta_2$,
whereas $\vartheta_1$ does not.
The labeling is identical to that of Table~\ref{tab:graphs-reduced-t3}.
\label{tab:g6}}
\end{table}

The central tool we use is state polynomial optimization
by Klep et al.~\cite{klep2023state}, also known as scalar extension~\cite{pozas2019, pozas2019quantum},
and which can be thought of as a variant of the non-commutative optimization framework
by Navascu\'es, Pironio, and Ac\'in~\cite{pironio2010}~\footnote{
While the scalar extension hierarchy was formulated earlier, Ref.~\cite{klep2023state} shows also completeness of the hierarchy.
We here use the term state polynomial optimization,
as it highlights the connections to polynomial optimization best.
In any case, the two terms can be used interchangeably.}.
%
%
This makes the resulting bounds both independent of the local dimension
and applicable to a wide range of situations.
Different normalization conditions (e.g. projectors, unitaries, unipotents),
commutation relations,
and additive as well as multiplicative uncertainty relations in higher order moments can be treated.
For example, Table~\ref{fig:higher_spin}
shows values of $\beta$ for a set of Heisenberg-Weyl operators
that satisfy commutation relations of the form
$A_i A_j = \zeta_{ij} A_j A_i$ with $\zeta_{ij}$ a root of unity.

In this Letter, we will restrict the exposition
to unitary operators and quadratic expressions as in Eq.~\eqref{eq:C}.
In what follows, upper case letters will refer to observables as in
$\langle A \rangle_\varrho := \tr (\varrho A)$,
while the lower case refers to
letters $a$ and words $w$ in the state polynomial framework,
whose evaluation on a state is referred to as $\langle a \rangle$
and $\langle w \rangle$.

\section{Lov\'{a}sz bound}
\label{sec:lovasz}
We introduce the upper bound on $\comp$ in Eq.~\eqref{eq:C} given by the Lov\'{a}sz theta number~\cite{de2022uncertainty}.
Let $\{A_1, \ldots, A_n\}$ be a collection of hermitian unitary operators that pairwise either commute or anti-commute,
for example, a collection of $n$-qubit Pauli operators.
We encode the commutation relations $A_i A_j = \zeta_{ij} A_j A_i$ with $\zeta_{ij}$
in a graph $G$ with adjacency matrix $\Gamma_{ij} = (1 - \zeta_{ij})/2$,
where $\zeta_{ij} \in \{+1, -1\}$.
In this way,
every vertex represents an operator, with two vertices connected whenever the corresponding operators anti-commute.
The Lov\'{a}sz bound then constrains $\beta$ as~\cite{de2022uncertainty, hastings2022optimizing}
\begin{equation}
 \comp \leq \vartheta(G)\,.
\end{equation}
Here $\vartheta(G)$ is the Lov\'{a}sz theta number of the graph $G$.

A for us convenient definition of $\vartheta(G)$ is through the optimal value of the following semidefinite program~\cite{GALLI2017159},
\begin{align}
\vartheta(G) \,=\, \max_{M} \quad & \sum_{i=1}^n M_{ii}
\label{eq:lovasz}\\
\textup{s.t.} \quad
         & M_{ii} = a_i, \nn \\
         & M_{ij} = 0 \quad \quad\quad\text{if } \zeta_{ij} = -1, \nn \\
         & \Delta = \begin{pmatrix}
         1 & a^T \\
         a & M
         \end{pmatrix}\geq 0. \nn
\end{align}

To see how Eq.~\eqref{eq:lovasz} gives an upper bound on Eq.~\eqref{eq:C},
that is  $\comp \leq \vartheta(G)$,
consider the following relaxation:
For any $\varrho$ construct a moment matrix~$\Gamma$ indexed by the set
$\{\one, \state{A_1^\dag} A_1, \dots, \state{A_n^\dag}A_n\}$
where $\state{A_i}=\tr(\varrho A_i)$ and $A_0 = \one$,
\begin{equation}
 \Gamma_{ij} = \tr\big(\varrho\, \state{A_i} \state{A_j^\dag} A_i^\dag A_j\big) \, .
\end{equation}
Such matrix $\Gamma$ has the form
\begin{align}
\kbordermatrix{
      & \one    & \state{A_1^\dag} A_1 & \cdots    & \state{A_n^\dag} A_n \\
 \one & 1       & \Gamma_{01}           &\cdots     & \Gamma_{0n} \\
 \state{A_1} A_1^\dag & & \Gamma_{11}    & \hdots   & \Gamma_{1n}\\
 \vdots &       &   & \ddots    & \vdots\\
 \state{A_n} A_n^\dag & &           &           & \Gamma_{nn}
  }\,.
  \label{eq:lovasz-moments}
\end{align}

In particular, $\Gamma_{0i} = \Gamma_{ii} = \state{A_i}^2$ since $A_i^\dag A_i = \one$,
and ${\Gamma}$ is positive semidefinite by construction.
Also, we can set $\Gamma_{ij} = 0$ when $\zeta_{ij} = -1$, because if $\Gamma$ is feasible, so is $(\Gamma + \Gamma^T)/2$
with the objective value unchanged.
Note that these last three properties coincide with the constraints imposed on $\Delta$ in Eq.~\eqref{eq:lovasz}.
However, a moment matrix~$\Gamma$ must additionally arise from some quantum state,
and generally the set of matrices $\Delta$ can be larger than the set of matrices $\Gamma$ arising from states.
Consequently, the optimum of Eq.~\eqref{eq:lovasz} upper bounds $\comp$ and the Lov\'{a}sz bound holds,
$
 \comp \leq \vartheta (G)
$.

While this approach already gives tight upper bounds in several settings, the pentagon graph in Table~\ref{tab:statepop-nh} with $\comp = 2 < \sqrt{5} = \vartheta(G)$
shows that this Lov\'{a}sz bound is not tight in general.
It turns out that a slight modification of the program in Eq.~\eqref{eq:lovasz}
can significantly strengthen the bounds.
The key idea is to consider larger moment matrices,
so to incorporate additional commutativity constraints
between products of operators.

\section{A Lov\'{a}sz-type hierarchy}
To obtain stronger bounds, consider the set of moment matrices indexed by
$\{\state{A_i^\dag} \state{A_j^\dag} A_i A_j\}_{0 \leq i < j \leq n}$ \footnote{Interestingly, we observe a similar numerical behaviour with the sequence $\{\state{(A_i A_j)^\dag} A_i A_j\}_{0 \leq i < j \leq n}$.},

\begin{align}
 \Gamma_{ij,kl} &= \tr\big(\varrho\, \state{A_i}\state{A_j}\state{A_k^\dag}\state{A_l^\dag} (A_i A_j)^\dag A_k A_l \big )\,.
\end{align}
These have the form
\begin{align}
\kbordermatrix{
      & \one    & \cdots   & \state{A_k^\dag}\state{A_l^\dag} A_k A_l & \cdots \\
 \one & 1       & \cdots     & \Gamma_{00,kl} & \cdots \\
 \vdots & & \ddots & \vdots  \\
 \state{A_i}\state{A_j} (A_i A_j)^\dag &       &   & \Gamma_{ij, kl}    & \hdots & \\
\vdots & &   &    & \ddots \\
}\, ,
  \label{eq:lovasz-moments-2}
\end{align}
The submatrix $\big(\Gamma_{i0,j0}\big)_{0 \leq i < j \leq n}$ coincides with Eq.~\eqref{eq:lovasz-moments},
while the complete matrix satisfies additional second-order commutation relations,
\begin{align}\label{eq:lovasz-R2}
         \Gamma_{ij,kl} = \zeta_{ji}  \Gamma_{\underline{ji}, kl} =
         \zeta_{kl}  \Gamma_{ij, \underline{lk}}
         = \zeta_{ik} \Gamma_{\underline{k}j,\underline{i}l}\,.
\end{align}
Again, $\Gamma$ is positive semidefinite.

Thus in analogy to Eq.~\eqref{eq:lovasz}, we formulate the relaxation
\begin{align}
\vartheta_{2}(G) \,=\,
\max_M \quad & \sum_{i=1}^n M_{i0,i0} \label{eq:lovasz2} \\
\text{s.t.} \quad
         &M_{ij,kl} = \zeta_{ji} M_{ji, kl}\,, \nn\\
         &M_{ij,kl} = \zeta_{kl} M_{ij, lk}\,, \nn \\
         &M_{ij,kl} = \zeta_{ik} M_{kj, il}\,, \nn \\
         &M_{ii,kl} = M_{0i, kl}\,, \nn \\     
         & M \geq 0\,. \nn
\end{align}
As in the previous section, we can restrict to real matrices:
if $M$ is feasible, then so is $(M + M^\dag)/2$ with unchanged objective value.
Thus we can set $M_{ij,kl} = 0$ if $(A_i A_j)^\dag A_k A_l = - (A_k A_l)^\dag A_i A_j$.
Later on, we cover the general case with non-hermitian operators and complex phases [Eq.~\eqref{eq:statepop-relax}], which requires complex matrices. 

In analogy to the argument made for the Lov\'{a}sz bound,
any $\Gamma$ matrix arising from a state also
satisfies the constraints imposed on the matrix $M$ in Eq.~\eqref{eq:lovasz2}.
Thus the set of valid moment matrices $\Gamma$ is contained in the set of feasible matrices~$M$.
Consequently, $\comp \leq \vartheta_{2}(G) $.
Additionally, as the submatrix $\big(M_{i0,j0}\big)_{i,j=0}^n$ coincides with $\Delta$ of Eq.~\eqref{eq:lovasz}, we have the strengthening
\begin{equation}
 \comp \leq \vartheta_{2}(G) \leq \vartheta(G)\,.
\end{equation}

For the pentagon graph in Table~\ref{tab:statepop-nh}, this enlarged program already yields a tight bound on $\comp = 2 = \vartheta_{2}(G)$.
For the remaining graphs with five vertices the Lov\'asz number equals the independence number,
and thus $\vartheta_{2}(G)$ is tight for all graphs with up to five vertices.

The argument above easily generalizes to larger indexing sequences that incorporate commutation relations between products of up to $k$ operators. Denote $\vec i = (i_1, \ldots, i_k)$,
a moment matrix $\Gamma$ indexed by
$\{ \state{A_{i_1}^\dag} \ldots \state{A_{i_k}^\dag} A_{i_1} \ldots A_{i_k}\}_{i_1, \ldots, i_k = 0}^n
$
where all non-zero indices are pairwise distinct and ordered increasingly. This
has entries 
\begin{align}
\Gamma_{\vec i, \vec j} = &\state{A_{i_1}}\ldots \state{A_{i_k}} \state{A_{j_1}^\dag} \dots \state{A_{j_k}^\dag} \nn\\
& \times (\state{A_{i_1} \ldots A_{i_k})^\dag A_{j_1} \ldots A_{j_k}}
\end{align}
satisfies the semidefinite program
\begin{align}
\vartheta_k(G) = \max_M \ \ & \sum_{i=1}^n M_{0\ldots 0 i,0\ldots 0i} \label{eq:lovaszk} \\
\text{s.t.} \ \
          & M_{\pi(a, b)(\vec i, \vec j)}  = \xi_{ab} M_{\vec i, \vec j} \, , \nn \\
         & M_{ii\ldots i_k, \vec j} = M_{0i \ldots i_k, \vec j}\, , \nn \\
         & M \geq 0 \, . \nn
\end{align}
Here, $\pi(a, b)$ is the transposition exchanging the $a$'th and $b$'th elements in the joint sequence $(i_k, \ldots, i_1, j_1, \ldots, j_k)$
(note the reversed indexing in the index $i$ arising from the dagger in the moment matrix)
and
\begin{equation}
\xi_{ab} = \prod_{a < c \leq b} \zeta_{ac} \zeta_{cb}\,
\end{equation}
is the factor that appears from the exchange.

By construction, Eq.~\eqref{eq:lovaszk} defines a hierarchy of upper bounds $\vartheta_k(G)$
\begin{equation}
    \comp(G) \leq \ldots \leq \vartheta_k(G) \leq  \ldots \leq \vartheta_1(G) = \vartheta(G) \, .
    \label{eq:sandwich2}
\end{equation}

We show in the next section that the relaxations $\vartheta_k$ in Eq.~\eqref{eq:lovaszk}, in particular $\vartheta$ in Eq.~\eqref{eq:lovasz}, can be enlarged to a complete hierarchy of semidefinite programs converging to the optimal value of Eq.~\eqref{eq:C}.

\section{A complete hierarchy}
\label{sec:optimization}

We now present a complete hierarchy that converges to~$\comp$.
Let $\zeta$ be a matrix 
encoding the commutation relations between a collection of operators $\{A_1, \ldots, A_n \}$.
The upper bound $\comp$ in Eq.~\eqref{eq:C}
can be formulated as the following optimization problem over expectations $\langle \,\cdot\, \rangle_\varrho$\,,
\begin{align}
\sup_\varrho \ \ & \sum_{i=1}^n|\state{A_i}|^2 \label{eq:statepop} \\
\textup{s.t.} \ \ &  A_i^\dag A_i = A_i A_i^\dag = 1\, , \nn \\
         & A_i A_j = \zeta_{ij} A_j A_i \, , \nn \\
         & \state{\one} = 1\, . \nn
\end{align}
We now show how to tackle this problem with a variant of   
non-commutative polynomial optimization~\cite{pironio2010}, allowing for non-linear expressions in the expectations, known as state polynomial optimization~\cite{klep2023state} or scalar extension~\cite{Tavakoli_2022}.

To solve this type of optimization problem,
consider words (or non-commutative monomials) $w = a_{j_1} \cdots a_{j_p}$ built from letters $\{a_i\}_{i=1}^n$
and expectation value of words denoted by $\prom{w} = \langle a_{j_1} \cdots a_{j_p} \rangle$.
The most general monomials in words and expectations are then of the form
$
 w = w_{0} \prom{w_1} \cdots \prom{w_m}
$.
Such formal state monomials can be added and multiplied to form state polynomials.
The involution of a word is defined by $(a_1 \ldots a_n)^* = a_n^* \ldots a_1^*$,
which plays the role of the adjoint of an operator in the formal setting.
In particular, $a_i^* = a_i$ when the operators are required to be hermitian.
Also, expectations behave as scalars, so that $v \prom{w} = \prom{w} v$ for all $v,w$.
Finally, $\prom{v\prom{w}} = \prom{\prom{v}\prom{w}} = \prom{v} \prom{w}$
and $(v\prom{w})^* = v^*\prom{w}^* = \prom{w}^* v^*$,
while $e$ is the identity (or empty) word satisfying $w e = w = e w$ for all words $w$.

For our problem, one additionally imposes the constraints $a_i a_j = \zeta_{ij} a_j a_i$
arising from the commutation relations $A_i A_j = \zeta_{ij}A_j A_i$.
Likewise, the constraint $A_i^\dag A_i = A_i A_i^\dag = \one$ gives rise to $a_i^* a_i = a_i a_i^* = e$.
For Pauli matrices, this reduces all monomials to square-free monomials.

Now consider a moment matrix $M_\ell$,
indexed by all state monomials of degree at most $\ell$,
whose entries are $M_\ell(v,w) = \langle v^*w\rangle$.
Then, the optimal solution of Eq.~\eqref{eq:statepop} is approximated from above with the following hierarchy of semidefinite programs with $\ell \in \N$~\cite[Lemma 6.5]{klep2023state},

\begin{align}
\Ibound(\zeta) & = \max \ \  \sum_{i=1}^n |\ss{a_i}|^2
\label{eq:statepop-relax} \\
\textup{s.t.} \ \ & \ss{\id} = 1\,,\nn \\
                  & M_\ell(v, w) = M_\ell(x, y) \quad \textup{when} \quad \ss{v^*w} = \ss{x^*y}\,, \nn \\
                  & M_\ell \geq 0\,, \nn
\end{align}
where one imposes on the entries $M_\ell$
all relations arising from $a_i a_j = \zeta_{ij} a_j a_i$
and $a^* a = aa^* = e$.
By increasing the degree $\ell$,
these relaxations converge to the optimal solution of Eq.~\eqref{eq:C}~\cite[Theorem 5.5 and Proposition 6.7]{klep2023state},
\begin{equation}
 \lim_{\ell \to \infty} \Ibound(\zeta) = \comp\,.
\end{equation}
For this result to hold, it is necessary that the optimization is over a bounded set of operators.
The technical condition is that the set is Archimedean,
that is,
there exists a constant $C>0$,
such that
$\sum_{i=1}^n a_ia_i^* \leq C$.
In our problem this is satisfied with $C=n$.
Additionally,
when the rank loop condition is met,
that is $\operatorname{rank}(M_\ell) = \operatorname{rank}(M_{\ell+1})$ (this is called a flat extension),
then the optimum has been reached~\cite[Proposition 6.10]{klep2023state}.

It can now be seen that $\vartheta_k(G)$ of Eq.~\eqref{eq:lovaszk} arises from the state polynomial optimization framework [Eq.~\eqref{eq:statepop-relax}]
where one considers only state monomials of the form
\begin{equation}
\{\ss{a_{i_1}^*} \cdots \ss{a_{i_k}^*} \, a_{i_1} \cdots a_{i_k} \}_{i_1, \ldots, i_k=0}^n\, ,
\label{eq:lovasz-index-k}
\end{equation}
with all non-zero indices pairwise distinct.

\section{Non-hermitian operators}

Note that both the complete and relaxed hierarchies can easily deal with operators that neither commute nor anti-commute,
but for which there is a complex phase $\zeta_{ij} \in \C$.
Then for every letter $a$ we define two new symbols to denote its real $\Re w$ and imaginary $\Im w$ part.
Additionally, for each word $w$ impose the constraints
$\Re w = \Re {w^*}$ and $\Im w = - \Im {w^*}$.
The positive semidefinite constraint on the complex moment matrix $M_\ell = \Re M_\ell + i \Im M_\ell \geq 0$
is equivalent to
\begin{equation}\label{eq:complex_moments}
\begin{pmatrix}
\Re M_\ell & - \Im M_\ell \\
\Im M_\ell & \phantom{-} \Re M_\ell
\end{pmatrix} \geq 0\,.
\end{equation}

Table~\ref{tab:statepop-nh} shows the first two levels of the hierarchy of Lov\'{a}sz relaxations for the Heisenberg-Weyl operators
computed through the complex to real mapping of Eq.~\eqref{eq:complex_moments}.
These operators are not hermitian in dimensions~$d$ greater than two,
and we use weighted graphs to specify their commutation relations where $\zeta_{ij}$ is a $d$'th root of unity.
Note that the cost of solving the problem in Eq.~\eqref{eq:statepop-relax} does not increase
with $d$ but only with the number of operators and the level of the hierarchy $k$.

\begin{table}[tbp]
\centering
\includegraphics{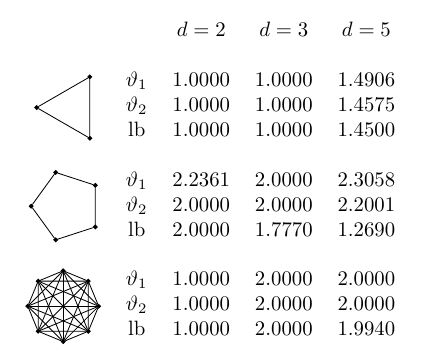}
\caption{\label{fig:higher_spin}
{\bf Upper bounds $\vartheta_k$ for higher dimensional spin operators},
obtained with an adaptation of Eq.~\eqref{eq:lovaszk} for commutation relations given by $d$'th roots of unity, $A_i A_j = e^{2\pi i/d} A_j A_i$ when $i \sim j$. This requires complex moment matrices that can be evaluated through Eq.~\eqref{eq:complex_moments}.
Lower bounds (lb) are obtained from sampling Haar random states and evaluating them on Heisenberg-Weyl operators [Eq.~\eqref{eq:heisenberg-weyl}].
Displacement operators [Eq.~\eqref{eq:displacement}] give the same bounds.
}
\label{tab:statepop-nh}
\end{table}

\section{Cutting planes}

\begin{table}[tbp]
\centering
\includegraphics{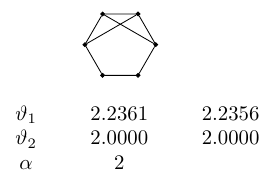}
\caption{\textbf{Cutting planes in graph $\# 3$.}
Left:  $\vartheta_k$ of Eq.~\eqref{eq:lovaszk}.
Right: $\vartheta_k$ strengthened with two intertwined 5-holes inequalities [Eq.~\eqref{eq:holes}].
\label{fig:c5-overlap}}
\end{table}

The Lov\'{a}sz bound in Eq.~\eqref{eq:lovasz} can be strengthened with additional constraints,
for instance with so-called odd-hole inequalities~\cite{GALLI2017159}.
Let $G$ be the anti-commutation graph of some hermitian unitary operators,
and let $H$ be a subset of vertices of $G$ inducing an odd cycle (i.e. an odd hole). Then
\begin{equation}
\sum_{i\in H} \state{A_i}^2 \leq \Big\lfloor \frac{|H|}{2} \Big\rfloor \, .
\label{eq:holes}
\end{equation}
This follows from Lemma $3$ of Ref.~\cite{PRXQuantum.5.020318}: Denote
by $C_l$ a cycle with $l$ vertices, then
$
 \beta(C_{2n+3}) - \beta(C_{2n+1}) \leq 1
$
for all $n\in \N$.
By Eq.~\eqref{eq:xyz} the cycle with three vertices is bounded by $\beta(C_3) \leq 1$
and Eq.~\eqref{eq:holes} follows by induction.

This constraint acts as a half-plane in the semidefinite program, and can strengthen the Lov\'asz bound for odd~$|H|$.
For graphs with up to seven vertices with non-overlapping holes,
these additional constraints are enough to tighten the bound for $\beta$.
However, these are not enough when the holes are intertwined as in Table~\ref{fig:c5-overlap}
(\#3 in Table~\ref{tab:graphs-reduced-t3}).
More generally,
one can impose constraints that arise
from higher levels in the hierarchy [Eq.~\eqref{eq:statepop-relax}]
on selected subgraphs.

\section{Conclusions}

We presented in Eq.~\eqref{eq:statepop-relax} a complete semidefinite programming hierarchy
converging to tight uncertainty relations.
The reduced formulation of this hierarchy in Eq.~\eqref{eq:lovaszk} can be seen as a natural generalization
of the Lov\'{a}sz bound ~\cite{de2022uncertainty}.
Interestingly, the second level of this reduced hierarchy already provides tight
bounds on $\comp$ for all graphs with no more than six vertices.
This answers an open question by Xu et al. on how to efficiently bound~$\beta$~\cite{PRXQuantum.5.020318}.
Additionally, our hierarchy applies to a wide range of scenarios, including
the $n$-qubit Pauli group,
generalised Pauli operators for higher-dimensional systems,
Clifford algebras,
and fermionic operators.
It also applies to higher order moments
and multiplicative uncertainty relations.
Several questions remain:
\begin{itemize}[noitemsep, topsep=0pt]
 \item Can a similar approach be derived for entropic uncertainty relations~\cite{PhysRevLett.60.1103, Wehner_2008, puchala2013majorization, korzekwa2014quantum, Zhao_2019, rotundo2023entropic}?

 \item Can uncertainty relations for Gell-Mann matrices
       be found that make use of the structure constants of $SU(3)$, in analogy to the Lov\'{a}sz bound?

 \item Is the hierarchy of relaxations $\vartheta_k(G)$ in Eq.~\eqref{eq:lovaszk} complete, in the sense that it converges to $\beta$?

 \item Can sparsity~\cite{wang2021exploiting} or symmetry~\cite{ioannou2021noncommutative, bachoc2012invariant, permenter2020dimension, brosch2022jordan, ligthart2022inflation} reductions boost efficiency of the SDP?

 \item Which hyperplane constraints improve the Lov\'asz bound most
        without incurring a too high computational overhead~\cite{GALLI2017159}?
        How do these constraints differ from the classical case?

 \item The weight enumerators describing quantum error correcting codes
        can be expressed as sums over $|\ss{A_i}|^2$.
       Can also here a Lov\'asz bound be established to upper bound the size of a quantum code?

 \item Can structure theorems for quasi-Clifford algebras or quantum tori~\cite{gastineau1982quasi, panov1998quantum} help to derive algebraic bounds?
 Are there any significant changes with respect to the choice of the matrix basis~\cite{Vourdas_2004},
 for example when considering nice error bases~\cite{klappenecker2001stabilizer}?

\end{itemize}

{\it Acknowledgements.---}
MBM and FH were supported by the Foundation for Polish Science through TEAM-NET (POIR.04.04.00-00-17C1/18-00).
We thank
Carlos de Gois,
Otfried G\"{u}hne,
Kiara Hansenne,
Igor Klep,
Victor Magron,
Stefano Pironio,
Alex Pozas,
Ren\'{e} Schwonnek,
Jurij Vol\v{c}i\v{c},
Zhen-Peng Xu,
and Karol \.Zyczkowski
for helpful comments and fruitful discussions.
We thank the organizers of the Quantum Corona 2022 workshop for hospitality and the friendly atmosphere.

\appendix

\section{Further applications}

Already for hermitian operators,
Eq.~\eqref{eq:C} finds applications in the characterization of entanglement and non-locality.
For example, Eq.~\eqref{eq:xyz} gives the entanglement witness~\cite{de2022uncertainty}
\begin{equation}
 W = \one \ot \one - \sigma_x\ot \sigma_x - \sigma_y \ot \sigma_y + \sigma_z \ot \sigma_z\,.
\end{equation}
It holds that $\prom{W}_{\varrho} \geq 0$ for all separable states $\varrho$,
whereas a joint eigenstate of $\sigma_x\otimes \sigma_x$, $\sigma_y\otimes \sigma_y$, and $\sigma_z\otimes \sigma_z$ achieves a value $\prom{W}_\varrho = -2$.

Similar constraints involving Pauli strings bound the quantum value of Bell inequalities~\cite{PhysRevLett.88.210401} and Bell monogamy~\cite{kurzynski2011correlation}. For instance, let $\LL$ be a Bell inequality with two measurements with two outcomes between two parties. Normalizing the value over classical correlations as $\LL \leq 1$, the optimal quantum value is obtained by the value in Eq.~\eqref{eq:C} for the following Pauli strings
\begin{equation}
\LL^2 \leq \ss{\sigma_x\otimes \sigma_x}^2 + \ss{\sigma_x\otimes \sigma_y}^2 + \ss{\sigma_y\otimes \sigma_x}^2 + \ss{\sigma_y\otimes \sigma_y}^2 \leq 2\,.
\label{eq:bell-bound}
\end{equation}
This gives the Tsirelson's bound for the CHSH inequality.
More generally, upper bounds on Eq.~\eqref{eq:C} for Pauli strings yield upper bounds for the quantum value of Bell inequalities and Bell monogamy involving only dichotomic measurements~\cite{PhysRevLett.88.210401, kurzynski2011correlation}. It is an interesting question whether the bound in Eq.~\eqref{eq:bell-bound} extends to measurements with more than two outcomes.

As noticed in Ref.~\cite{PRXQuantum.5.020318}, Eq.~\eqref{eq:C} also bounds the ground state energy of the Hamiltonian $H = \sum_i^n A_i$, since $\prom H^2 \leq n \sum_i^n \prom{A_i}^2$. These bounds are determined by the algebraic relations between the terms and thus are independent of the physical dimension.

\section{The Lov\'{a}sz theta number}

A \emph{graph} $G = (V, E)$ consists of a set $V$ of \emph{vertices} and a set $E \subset V \times V$ of \emph{edges} between them. Two vertices $u, v \in V$ are connected ($u\sim v$) when $(u, v)\in E$,
and disconnected or \emph{independent} otherwise.
The \emph{complement} of $G$ is $\bar G = (V, \bar E)$, where $(u, v) \in \bar E$ if and only if $(u, v) \not \in E$.

The \emph{independence number}~$\alpha(G)$ is the maximum number of pairwise independent vertices of $G$.
The \emph{chromatic number}~$\chi(G)$ is the minimum number of different colors needed to assign to the vertices of $G$, such that no connected vertices share the same color.
These quantities, also called \emph{graph invariants}, encode key combinatorial properties of $G$ and are NP-complete \cite{Karp1972}.

However, some graph invariants capture properties of the graph that are easier to compute.
Of particular interest is the \emph{Lov\'{a}sz number} $\vartheta(G)$,
which is sandwiched by the two quantities described above~\cite{knuth1993sandwich},
\begin{equation}
\alpha (G) \leq \vartheta (G) \leq \chi (\bar G) \, .
\end{equation}
It can be computed in polynomial time since
it can be defined through a semidefinite program~\cite{lovasz1979},
\begin{align}
\vartheta(G) \,=\, \max_{M} \quad & \sum_{i,j=1}^n M_{ij}
\label{eq:lovasz1}\\
\textup{s.t.} \quad
         & \tr(M) = 1 \nn \\
         & M_{ij} = 0 \quad \quad\quad\text{if } i \sim j \nn \\
         & M \geq 0\,. \nn
\end{align}
Equivalent definitions for $\vartheta (G)$ are~\cite{GALLI2017159},
\begin{align}
\vartheta(G) \,=\, \max_{M} \quad & \sum_{i=1}^n M_{ii}\\
\textup{s.t.} \quad
         & M_{ii} = a_i \nn \\
         & M_{ij} = 0 \quad \quad\quad\text{if } i \sim j \nn \\
         & \Delta = \begin{pmatrix}
         1 & a^T \\
         a & M
         \end{pmatrix}\geq 0\,, \nn
\end{align}
and~\cite{knuth1993sandwich}
\begin{align} \label{eq:lovasz3}
\vartheta(G) \,=\, \max_{M} \quad &\lambda_{\max}(M)
\\
\textup{s.t.} \quad
         & M_{ii} = 1 \nn \\
         & M_{ij} = 0 \quad \quad\quad\text{if } i \sim j \nn \\
         & M \geq 0\,, \nn
\end{align}
where $\lambda_{\max}(M)$ denotes the maximum eigenvalue of $M$.

Given a collection of hermitian unitary operators $A_1, \ldots, A_n$ that mutually either commute or anti-commute,
define the \emph{anti-commutativity graph} $G$ on vertices $\{1, \ldots, n\}$ where $i \sim j$ if $A_i A_j = - A_j A_i$.
The quantitiy
\begin{equation}
 \comp =\sup_\varrho \sum_{i=1}^n \prom{A_i}_\varrho^2\,.
\end{equation}
is then a graph invariant that is defined solely through the commutation relations encoded in $G$.
Commuting operators can be jointly diagonalized, thus it is clear that $\alpha(G) \leq \comp(G)$: a joint eigenstate simultaneously achieves the value $\prom{A_i}_\varrho^2 = 1$ for each term in an independent set.
On the other hand, Ref.~\cite{de2022uncertainty} showed that $\comp (G) \leq \vartheta(G)$ by using the definition in Eq.~\eqref{eq:lovasz3}.

As an example, take the following five two-qubit Pauli observables
\begin{equation}
\sigma_x\otimes \sigma_x \, , \ 
\sigma_x\otimes \sigma_y \, , \ 
\id \otimes \sigma_x \, , \ 
\sigma_y \otimes \sigma_z \,, \
\sigma_y \otimes \sigma_x  \, .
\end{equation}
Their anti-commutativity graph is the pentagon $G = C_5$ shown in Table~\ref{tab:graphs-reduced-t3}. The second level of the hierarchy improves on the Lov\'{a}sz bound, closing the gap with the independence number and thus giving the tight bound
\begin{equation}
2 = \alpha(C_5) = \beta (C_5) = \vartheta_2(C_5) < \vartheta_1(C_5) = \sqrt{5}\, .
\end{equation}

\section{Comparison with previous work}

Upper bounds in Ref.~\cite{PRXQuantum.5.020318} are obtained fixing a representation of the operators $(A_1, \ldots, A_n)$ as Pauli strings in dimension $2^N$, through the separability problem
\begin{align}
\max_\sigma \quad & \sum_{i=1}^n\prom{A_i \otimes A_i}_\sigma \\
\text{s.t.} \quad & \sigma \in \operatorname{SEP} \nn \,.
\end{align}
Here the maximization runs over separable states on ${\C^2}^{\otimes N} \ot {\C^2}^{\otimes N}$.
This problem can be outer-approximated with semidefinite programs based on symmetric extension.
Level $k \in \N$ of this hierarchy corresponds to

\begin{align}
\max_\sigma \quad & \sum_{i=1}^n\prom{A_i \otimes A_i}_{\sigma} \\
\text{s.t.} \quad & \sigma \in \operatorname{Sym}_k \, .\nn
\end{align}
Here, the maximization runs over all states $\sigma=\sigma_{AB_1\dots B_k}$ on ${(\C^2)}^{\otimes N} \ot {(\C^2)}^{\otimes (N(1+k))}$
that are invariant under exchange of the subsystems $B_1,\dots, B_k$, with positive partial transpose across all bipartitions.
The dimension of this hierarchy grows as $2^{N(2+k)}$, which renders the relaxations intractable quickly.

For instance, take the graph $G = \bar C_7$, the complement of the cycle with seven vertices.
These operators can be represented with Pauli strings of $N=3$ qubits.
The first extension has size $2^9 = 512$, which is already quite large for a desktop computer to solve.

Our approach based on state polynomial optimization [Eq.~\eqref{eq:statepop}] does not fix a representation,
and is thus dimension-independent.
The size of the relaxations is given by the size of the indexing sequence for the moment matrix in Eq.~\eqref{eq:statepop-relax},
which allows a finer control and flexibility in the size.
In particular,
our hierarchy $\vartheta_k(G)$ in Eq.~\eqref{eq:lovaszk}
requires for a graph $G$ with $n$ vertices
to solve SDPs of size
$1+n$ for $\vartheta_1(G)$ [Eq.~\eqref{eq:lovasz}],
size $1+ n(n+1)/2$ for $\vartheta_2(G)$ [Eq.~\eqref{eq:lovasz2}],
and size $2^n + 1$ for $\vartheta_n (G)$.
For a graph $G$ with seven vertices such as $\bar C_7$,
$\vartheta_7(G)$ has size $2^7 = 128$,
which can easily be solved on a desktop computer.

In the approach by Ref.~\cite{PRXQuantum.5.020318}
the size of the resulting SDPs depends
on whether the commutation relations can be represented on a small number of qubits,
while it only depends on the number of observables for our hierarchies.
However, with the symmetric extension approach
one in principle is able to give approximation guarantees
through the quantum de Finetti theorem~\cite{Christandl2007}.

\section{Small graphs}

Table~\ref{tab:graphs-reduced-t3} shows
all 43 non-isomorphic graphs with up to seven vertices for which $\comp < \vartheta_1$.
Strengthening $\vartheta_1$ with odd-hole inequalities as in Eq.~\eqref{eq:holes},
we obtain tight bounds on $\beta$ for 18 graphs by matching the upper bounds with the independence number.
For all but $10$ of these graphs it holds that $\comp = \vartheta_2$,
and $\comp=\vartheta_3$ holds for all but $7$ graphs.

Our hierarchy also improves on the best known upper bounds for all $36$ non-isomorphic graphs with eight vertices and all $1256$ non-isomorphic graphs with nine vertices for which the value of $\beta$ is unknown~\cite{private_communication}.
Already our second level [Eq.~\eqref{eq:lovasz2}] gives tight bounds $\beta(G) = \vartheta_2(G)$ for four of those graphs, up to numerical precision.

\begin{table*}
\includegraphics[width=\textwidth]{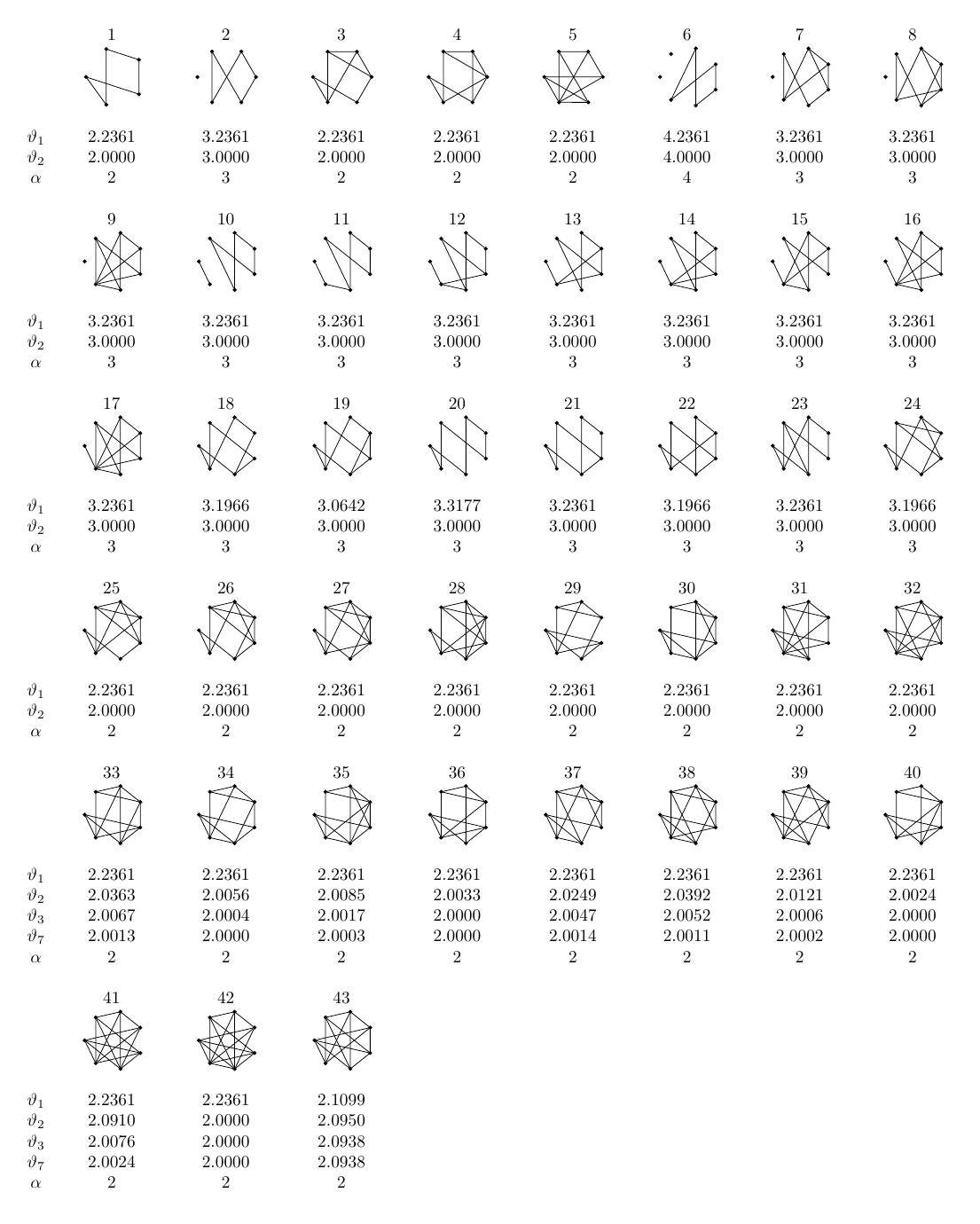}
\caption{\textbf{Hierarchy of upper bounds} for all non-isomorphic graphs up to seven vertices for which $\comp < \vartheta_1$.
Our hierarchy $\vartheta_k$ in Eq.~\eqref{eq:lovaszk} improves on the bounds given by the Lov\'{a}sz number $\vartheta_1$ in Eq.~\eqref{eq:lovasz} and closes the gap for all but 7 graphs: 33, 35, 37, 38, 39, 41 and 43.
\label{tab:graphs-reduced-t3}
}
\end{table*}

\section{Non-hermitian operators}
\label{sec:non-hermitian}

\noindent\textbf{Heisenberg-Weyl basis.}
The Heisenberg-Weyl operators can be seen as a generalization of the Pauli matrices.
They play a central role in the description of $d$-dimensional quantum spin systems~\cite{Vourdas_2004}
and in the construction of non-binary quantum codes~\cite{ketkar2006nonbinary}.
On the space $\C^d$ define the operators $X$ and $Z$ through~\cite{scott2004multipartite}
\begin{equation}
X \ket j = \ket {j+1}\,,  \quad Z \ket j = \omega^j \ket{j}\,,
\end{equation}
where $\omega = \exp(2\pi i /d)$ is the principal root of unity and the addition is taken modulo $d$.
These operators are related by the quantum Fourier transform in dimension $d$.
The collection of operators
\begin{equation}
\sigma(k, l) \ket i = X^k Z^l \ket i = \omega^{il} \ket{i+k}\,,
\label{eq:heisenberg-weyl}
\end{equation}
forms the Heisenberg-Weyl basis for the space of complex $d\times d$ matrices.
These satisfy the commutativity relations
\begin{equation}
\sigma(k, l) \sigma(m, n) = \omega^{lm - kn} \sigma(m, n) \sigma(k, l) \,.
\label{eq:heisenberg-weyl-com}
\end{equation}
Their adjoints are given by
\begin{equation}
\sigma(k, l)^\dag = \omega^{kl} \sigma(d-k, d-l)\,,
\label{eq:heisenber-weyl-adj}
\end{equation}
and they form an orthogonal basis
\begin{equation}
 \tr[\sigma(k, l)^\dag \sigma(m, n)] = \delta_{km}\delta_{ln} d\,.
\end{equation}

\noindent\textbf{Displacement basis.}
Introducing a phase~\cite{scott2004multipartite},
\begin{equation}
D(k, l) = \omega^{kl/2} \sigma(k, l)\,,
\label{eq:displacement}
\end{equation}
the adjoint operators are given by
\begin{equation}
D(k, l)^\dag = (-1)^{k + l + d} D(d-k, d-l)\,.
\end{equation}
These are called displacement operators,
and the algebra they generate is isomorphic to the one generated by Heisenberg-Weyl operators.
Thus, the bounds
for the non-hermitian version of Eq.~\eqref{eq:statepop} coincide for both choices of generators (c.f. Table~\ref{tab:statepop-nh}).


\begin{thebibliography}{51}%
\makeatletter
\providecommand \@ifxundefined [1]{%
 \@ifx{#1\undefined}
}%
\providecommand \@ifnum [1]{%
 \ifnum #1\expandafter \@firstoftwo
 \else \expandafter \@secondoftwo
 \fi
}%
\providecommand \@ifx [1]{%
 \ifx #1\expandafter \@firstoftwo
 \else \expandafter \@secondoftwo
 \fi
}%
\providecommand \natexlab [1]{#1}%
\providecommand \enquote  [1]{``#1''}%
\providecommand \bibnamefont  [1]{#1}%
\providecommand \bibfnamefont [1]{#1}%
\providecommand \citenamefont [1]{#1}%
\providecommand \href@noop [0]{\@secondoftwo}%
\providecommand \href [0]{\begingroup \@sanitize@url \@href}%
\providecommand \@href[1]{\@@startlink{#1}\@@href}%
\providecommand \@@href[1]{\endgroup#1\@@endlink}%
\providecommand \@sanitize@url [0]{\catcode `\\12\catcode `\$12\catcode
  `\&12\catcode `\#12\catcode `\^12\catcode `\_12\catcode `\%12\relax}%
\providecommand \@@startlink[1]{}%
\providecommand \@@endlink[0]{}%
\providecommand \url  [0]{\begingroup\@sanitize@url \@url }%
\providecommand \@url [1]{\endgroup\@href {#1}{\urlprefix }}%
\providecommand \urlprefix  [0]{URL }%
\providecommand \Eprint [0]{\href }%
\providecommand \doibase [0]{http://dx.doi.org/}%
\providecommand \selectlanguage [0]{\@gobble}%
\providecommand \bibinfo  [0]{\@secondoftwo}%
\providecommand \bibfield  [0]{\@secondoftwo}%
\providecommand \translation [1]{[#1]}%
\providecommand \BibitemOpen [0]{}%
\providecommand \bibitemStop [0]{}%
\providecommand \bibitemNoStop [0]{.\EOS\space}%
\providecommand \EOS [0]{\spacefactor3000\relax}%
\providecommand \BibitemShut  [1]{\csname bibitem#1\endcsname}%
\let\auto@bib@innerbib\@empty
\bibitem [{\citenamefont {Fuchs}\ and\ \citenamefont
  {Peres}(1996)}]{fuchs1996quantum}%
  \BibitemOpen
  \bibfield  {author} {\bibinfo {author} {\bibfnamefont {C.~A.}\ \bibnamefont
  {Fuchs}}\ and\ \bibinfo {author} {\bibfnamefont {A.}~\bibnamefont {Peres}},\
  }\href {\doibase https://doi.org/10.1103/PhysRevA.53.2038} {\bibfield
  {journal} {\bibinfo  {journal} {Phys. Rev. A}\ }\textbf {\bibinfo {volume}
  {53}},\ \bibinfo {pages} {2038} (\bibinfo {year} {1996})}\BibitemShut
  {NoStop}%
\bibitem [{\citenamefont {G{\"u}hne}(2004)}]{guhne2004characterizing}%
  \BibitemOpen
  \bibfield  {author} {\bibinfo {author} {\bibfnamefont {O.}~\bibnamefont
  {G{\"u}hne}},\ }\href {\doibase
  https://doi.org/10.1103/PhysRevLett.92.117903} {\bibfield  {journal}
  {\bibinfo  {journal} {Phys. Rev. Lett.}\ }\textbf {\bibinfo {volume} {92}},\
  \bibinfo {pages} {117903} (\bibinfo {year} {2004})}\BibitemShut {NoStop}%
\bibitem [{\citenamefont {Cavalcanti}\ and\ \citenamefont
  {Reid}(2007)}]{cavalcanti2007uncertainty}%
  \BibitemOpen
  \bibfield  {author} {\bibinfo {author} {\bibfnamefont {E.~G.}\ \bibnamefont
  {Cavalcanti}}\ and\ \bibinfo {author} {\bibfnamefont {M.~D.}\ \bibnamefont
  {Reid}},\ }\href {\doibase https://doi.org/10.1080/09500340701639623}
  {\bibfield  {journal} {\bibinfo  {journal} {J. Mod. Opt.}\ }\textbf {\bibinfo
  {volume} {54}},\ \bibinfo {pages} {2373} (\bibinfo {year}
  {2007})}\BibitemShut {NoStop}%
\bibitem [{\citenamefont {Renes}\ and\ \citenamefont
  {Boileau}(2009)}]{renes2009conjectured}%
  \BibitemOpen
  \bibfield  {author} {\bibinfo {author} {\bibfnamefont {J.~M.}\ \bibnamefont
  {Renes}}\ and\ \bibinfo {author} {\bibfnamefont {J.-C.}\ \bibnamefont
  {Boileau}},\ }\href {\doibase https://doi.org/10.1103/PhysRevLett.103.020402}
  {\bibfield  {journal} {\bibinfo  {journal} {Phys. Rev. Lett.}\ }\textbf
  {\bibinfo {volume} {103}},\ \bibinfo {pages} {020402} (\bibinfo {year}
  {2009})}\BibitemShut {NoStop}%
\bibitem [{\citenamefont {Gottesman}(2010)}]{gottesman2010introduction}%
  \BibitemOpen
  \bibfield  {author} {\bibinfo {author} {\bibfnamefont {D.}~\bibnamefont
  {Gottesman}},\ }in\ \href {\doibase https://doi.org/10.48550/arXiv.0904.2557}
  {\emph {\bibinfo {booktitle} {Quantum information science and its
  contributions to mathematics, Proceedings of Symposia in Applied
  Mathematics}}},\ Vol.~\bibinfo {volume} {68}\ (\bibinfo {year} {2010})\ pp.\
  \bibinfo {pages} {13--58}\BibitemShut {NoStop}%
\bibitem [{\citenamefont {Bennett}\ and\ \citenamefont
  {Brassard}(2014)}]{bennett2014quantum}%
  \BibitemOpen
  \bibfield  {author} {\bibinfo {author} {\bibfnamefont {C.~H.}\ \bibnamefont
  {Bennett}}\ and\ \bibinfo {author} {\bibfnamefont {G.}~\bibnamefont
  {Brassard}},\ }\href {\doibase https://doi.org/10.1016/j.tcs.2014.05.025}
  {\bibfield  {journal} {\bibinfo  {journal} {Theor. Comput. Sci.}\ }\textbf
  {\bibinfo {volume} {560}},\ \bibinfo {pages} {7} (\bibinfo {year}
  {2014})}\BibitemShut {NoStop}%
\bibitem [{\citenamefont {S\o{}rensen}\ and\ \citenamefont
  {M\o{}lmer}(2001)}]{sorensen2001}%
  \BibitemOpen
  \bibfield  {author} {\bibinfo {author} {\bibfnamefont {A.~S.}\ \bibnamefont
  {S\o{}rensen}}\ and\ \bibinfo {author} {\bibfnamefont {K.}~\bibnamefont
  {M\o{}lmer}},\ }\href {\doibase 10.1103/PhysRevLett.86.4431} {\bibfield
  {journal} {\bibinfo  {journal} {Phys. Rev. Lett.}\ }\textbf {\bibinfo
  {volume} {86}},\ \bibinfo {pages} {4431} (\bibinfo {year}
  {2001})}\BibitemShut {NoStop}%
\bibitem [{\citenamefont {T\'oth}\ and\ \citenamefont
  {Fr\"owis}(2022)}]{toth2022}%
  \BibitemOpen
  \bibfield  {author} {\bibinfo {author} {\bibfnamefont {G.}~\bibnamefont
  {T\'oth}}\ and\ \bibinfo {author} {\bibfnamefont {F.}~\bibnamefont
  {Fr\"owis}},\ }\href {\doibase 10.1103/PhysRevResearch.4.013075} {\bibfield
  {journal} {\bibinfo  {journal} {Phys. Rev. Res.}\ }\textbf {\bibinfo {volume}
  {4}},\ \bibinfo {pages} {013075} (\bibinfo {year} {2022})}\BibitemShut
  {NoStop}%
\bibitem [{\citenamefont {Dammeier}\ \emph {et~al.}(2015)\citenamefont
  {Dammeier}, \citenamefont {Schwonnek},\ and\ \citenamefont
  {Werner}}]{dammeier2015}%
  \BibitemOpen
  \bibfield  {author} {\bibinfo {author} {\bibfnamefont {L.}~\bibnamefont
  {Dammeier}}, \bibinfo {author} {\bibfnamefont {R.}~\bibnamefont {Schwonnek}},
  \ and\ \bibinfo {author} {\bibfnamefont {R.~F.}\ \bibnamefont {Werner}},\
  }\href {\doibase 10.1088/1367-2630/17/9/093046} {\bibfield  {journal}
  {\bibinfo  {journal} {New J. Phys.}\ }\textbf {\bibinfo {volume} {17}},\
  \bibinfo {pages} {093046} (\bibinfo {year} {2015})}\BibitemShut {NoStop}%
\bibitem [{\citenamefont {Szyma{\'n}ski}\ and\ \citenamefont
  {{\.Z}yczkowski}(2019)}]{szymanski2019}%
  \BibitemOpen
  \bibfield  {author} {\bibinfo {author} {\bibfnamefont {K.}~\bibnamefont
  {Szyma{\'n}ski}}\ and\ \bibinfo {author} {\bibfnamefont {K.}~\bibnamefont
  {{\.Z}yczkowski}},\ }\href {\doibase 10.1088/1751-8121/ab4543} {\bibfield
  {journal} {\bibinfo  {journal} {J. Phys. A Math. Theor.}\ }\textbf {\bibinfo
  {volume} {53}},\ \bibinfo {pages} {015302} (\bibinfo {year}
  {2019})}\BibitemShut {NoStop}%
\bibitem [{\citenamefont {de~Gois}\ \emph {et~al.}(2023)\citenamefont
  {de~Gois}, \citenamefont {Hansenne},\ and\ \citenamefont
  {G\"uhne}}]{de2022uncertainty}%
  \BibitemOpen
  \bibfield  {author} {\bibinfo {author} {\bibfnamefont {C.}~\bibnamefont
  {de~Gois}}, \bibinfo {author} {\bibfnamefont {K.}~\bibnamefont {Hansenne}}, \
  and\ \bibinfo {author} {\bibfnamefont {O.}~\bibnamefont {G\"uhne}},\ }\href
  {\doibase 10.1103/PhysRevA.107.062211} {\bibfield  {journal} {\bibinfo
  {journal} {Phys. Rev. A}\ }\textbf {\bibinfo {volume} {107}},\ \bibinfo
  {pages} {062211} (\bibinfo {year} {2023})}\BibitemShut {NoStop}%
\bibitem [{\citenamefont {Kurzy{\'n}ski}\ \emph {et~al.}(2011)\citenamefont
  {Kurzy{\'n}ski}, \citenamefont {Paterek}, \citenamefont {Ramanathan},
  \citenamefont {Laskowski},\ and\ \citenamefont
  {Kaszlikowski}}]{kurzynski2011correlation}%
  \BibitemOpen
  \bibfield  {author} {\bibinfo {author} {\bibfnamefont {P.}~\bibnamefont
  {Kurzy{\'n}ski}}, \bibinfo {author} {\bibfnamefont {T.}~\bibnamefont
  {Paterek}}, \bibinfo {author} {\bibfnamefont {R.}~\bibnamefont {Ramanathan}},
  \bibinfo {author} {\bibfnamefont {W.}~\bibnamefont {Laskowski}}, \ and\
  \bibinfo {author} {\bibfnamefont {D.}~\bibnamefont {Kaszlikowski}},\ }\href
  {\doibase https://doi.org/10.1103/PhysRevLett.106.180402} {\bibfield
  {journal} {\bibinfo  {journal} {Phys. Rev. Lett.}\ }\textbf {\bibinfo
  {volume} {106}},\ \bibinfo {pages} {180402} (\bibinfo {year}
  {2011})}\BibitemShut {NoStop}%
\bibitem [{\citenamefont {Loulidi}\ and\ \citenamefont
  {Nechita}(2022)}]{loulidi2022}%
  \BibitemOpen
  \bibfield  {author} {\bibinfo {author} {\bibfnamefont {F.}~\bibnamefont
  {Loulidi}}\ and\ \bibinfo {author} {\bibfnamefont {I.}~\bibnamefont
  {Nechita}},\ }\href {\doibase https://doi.org/10.1103/PRXQuantum.3.040325}
  {\bibfield  {journal} {\bibinfo  {journal} {PRX Quantum}\ }\textbf {\bibinfo
  {volume} {3}},\ \bibinfo {pages} {040325} (\bibinfo {year}
  {2022})}\BibitemShut {NoStop}%
\bibitem [{\citenamefont {Xu}\ \emph {et~al.}(2024)\citenamefont {Xu},
  \citenamefont {Schwonnek},\ and\ \citenamefont
  {Winter}}]{PRXQuantum.5.020318}%
  \BibitemOpen
  \bibfield  {author} {\bibinfo {author} {\bibfnamefont {Z.-P.}\ \bibnamefont
  {Xu}}, \bibinfo {author} {\bibfnamefont {R.}~\bibnamefont {Schwonnek}}, \
  and\ \bibinfo {author} {\bibfnamefont {A.}~\bibnamefont {Winter}},\ }\href
  {\doibase 10.1103/PRXQuantum.5.020318} {\bibfield  {journal} {\bibinfo
  {journal} {PRX Quantum}\ }\textbf {\bibinfo {volume} {5}},\ \bibinfo {pages}
  {020318} (\bibinfo {year} {2024})}\BibitemShut {NoStop}%
\bibitem [{\citenamefont {Hastings}\ and\ \citenamefont
  {O'Donnell}(2022)}]{hastings2022optimizing}%
  \BibitemOpen
  \bibfield  {author} {\bibinfo {author} {\bibfnamefont {M.~B.}\ \bibnamefont
  {Hastings}}\ and\ \bibinfo {author} {\bibfnamefont {R.}~\bibnamefont
  {O'Donnell}},\ }in\ \href {\doibase https://doi.org/10.1145/3519935.3519960}
  {\emph {\bibinfo {booktitle} {Proceedings of the 54th Annual ACM SIGACT
  Symposium on Theory of Computing}}}\ (\bibinfo {year} {2022})\ pp.\ \bibinfo
  {pages} {776--789}\BibitemShut {NoStop}%
\bibitem [{\citenamefont {Cabello}\ \emph {et~al.}(2014)\citenamefont
  {Cabello}, \citenamefont {Severini},\ and\ \citenamefont
  {Winter}}]{Cabello_2014}%
  \BibitemOpen
  \bibfield  {author} {\bibinfo {author} {\bibfnamefont {A.}~\bibnamefont
  {Cabello}}, \bibinfo {author} {\bibfnamefont {S.}~\bibnamefont {Severini}}, \
  and\ \bibinfo {author} {\bibfnamefont {A.}~\bibnamefont {Winter}},\ }\href
  {\doibase 10.1103/PhysRevLett.112.040401} {\bibfield  {journal} {\bibinfo
  {journal} {Phys. Rev. Lett.}\ }\textbf {\bibinfo {volume} {112}},\ \bibinfo
  {pages} {040401} (\bibinfo {year} {2014})}\BibitemShut {NoStop}%
\bibitem [{\citenamefont {Ac{\'\i}n}\ \emph {et~al.}(2015)\citenamefont
  {Ac{\'\i}n}, \citenamefont {Fritz}, \citenamefont {Leverrier},\ and\
  \citenamefont {Sainz}}]{acin2015combinatorial}%
  \BibitemOpen
  \bibfield  {author} {\bibinfo {author} {\bibfnamefont {A.}~\bibnamefont
  {Ac{\'\i}n}}, \bibinfo {author} {\bibfnamefont {T.}~\bibnamefont {Fritz}},
  \bibinfo {author} {\bibfnamefont {A.}~\bibnamefont {Leverrier}}, \ and\
  \bibinfo {author} {\bibfnamefont {A.~B.}\ \bibnamefont {Sainz}},\ }\href
  {\doibase https://doi.org/10.1007/s00220-014-2260-1} {\bibfield  {journal}
  {\bibinfo  {journal} {Commun. Math. Phys.}\ }\textbf {\bibinfo {volume}
  {334}},\ \bibinfo {pages} {533} (\bibinfo {year} {2015})}\BibitemShut
  {NoStop}%
\bibitem [{\citenamefont {Ac{\'{\i}}n}\ \emph {et~al.}(2017)\citenamefont
  {Ac{\'{\i}}n}, \citenamefont {Duan}, \citenamefont {Roberson}, \citenamefont
  {Sainz},\ and\ \citenamefont {Winter}}]{Acin_2017}%
  \BibitemOpen
  \bibfield  {author} {\bibinfo {author} {\bibfnamefont {A.}~\bibnamefont
  {Ac{\'{\i}}n}}, \bibinfo {author} {\bibfnamefont {R.}~\bibnamefont {Duan}},
  \bibinfo {author} {\bibfnamefont {D.~E.}\ \bibnamefont {Roberson}}, \bibinfo
  {author} {\bibfnamefont {A.~B.}\ \bibnamefont {Sainz}}, \ and\ \bibinfo
  {author} {\bibfnamefont {A.}~\bibnamefont {Winter}},\ }\href {\doibase
  10.1016/j.dam.2016.04.028} {\bibfield  {journal} {\bibinfo  {journal}
  {Discrete Appl. Math.}\ }\textbf {\bibinfo {volume} {216}},\ \bibinfo {pages}
  {489} (\bibinfo {year} {2017})}\BibitemShut {NoStop}%
\bibitem [{\citenamefont {Duan}\ \emph {et~al.}(2013)\citenamefont {Duan},
  \citenamefont {Severini},\ and\ \citenamefont {Winter}}]{duan2013zero}%
  \BibitemOpen
  \bibfield  {author} {\bibinfo {author} {\bibfnamefont {R.}~\bibnamefont
  {Duan}}, \bibinfo {author} {\bibfnamefont {S.}~\bibnamefont {Severini}}, \
  and\ \bibinfo {author} {\bibfnamefont {A.}~\bibnamefont {Winter}},\ }\href
  {\doibase 10.1109/TIT.2012.2221677} {\bibfield  {journal} {\bibinfo
  {journal} {IEEE Trans. Inf. Theory}\ }\textbf {\bibinfo {volume} {59}},\
  \bibinfo {pages} {1164} (\bibinfo {year} {2013})}\BibitemShut {NoStop}%
\bibitem [{pri()}]{private_communication}%
  \BibitemOpen
  \href@noop {} {}\bibinfo {note} {Private communication with the authors of
  Ref.~\cite{PRXQuantum.5.020318}}\BibitemShut {NoStop}%
\bibitem [{\citenamefont {Klep}\ \emph {et~al.}(2023)\citenamefont {Klep},
  \citenamefont {Magron}, \citenamefont {Vol{\v{c}}i{\v{c}}},\ and\
  \citenamefont {Wang}}]{klep2023state}%
  \BibitemOpen
  \bibfield  {author} {\bibinfo {author} {\bibfnamefont {I.}~\bibnamefont
  {Klep}}, \bibinfo {author} {\bibfnamefont {V.}~\bibnamefont {Magron}},
  \bibinfo {author} {\bibfnamefont {J.}~\bibnamefont {Vol{\v{c}}i{\v{c}}}}, \
  and\ \bibinfo {author} {\bibfnamefont {J.}~\bibnamefont {Wang}},\ }\href@noop
  {} {\enquote {\bibinfo {title} {State polynomials: positivity, optimization
  and nonlinear {B}ell inequalities},}\ } (\bibinfo {year} {2023}),\ \Eprint
  {http://arxiv.org/abs/2301.12513} {arXiv:2301.12513} \BibitemShut {NoStop}%
\bibitem [{\citenamefont {Pozas-Kerstjens}\ \emph {et~al.}(2019)\citenamefont
  {Pozas-Kerstjens}, \citenamefont {Rabelo}, \citenamefont {Rudnicki},
  \citenamefont {Chaves}, \citenamefont {Cavalcanti}, \citenamefont
  {Navascu\'es},\ and\ \citenamefont {Ac\'{\i}n}}]{pozas2019}%
  \BibitemOpen
  \bibfield  {author} {\bibinfo {author} {\bibfnamefont {A.}~\bibnamefont
  {Pozas-Kerstjens}}, \bibinfo {author} {\bibfnamefont {R.}~\bibnamefont
  {Rabelo}}, \bibinfo {author} {\bibfnamefont {L.}~\bibnamefont {Rudnicki}},
  \bibinfo {author} {\bibfnamefont {R.}~\bibnamefont {Chaves}}, \bibinfo
  {author} {\bibfnamefont {D.}~\bibnamefont {Cavalcanti}}, \bibinfo {author}
  {\bibfnamefont {M.}~\bibnamefont {Navascu\'es}}, \ and\ \bibinfo {author}
  {\bibfnamefont {A.}~\bibnamefont {Ac\'{\i}n}},\ }\href {\doibase
  10.1103/PhysRevLett.123.140503} {\bibfield  {journal} {\bibinfo  {journal}
  {Phys. Rev. Lett.}\ }\textbf {\bibinfo {volume} {123}},\ \bibinfo {pages}
  {140503} (\bibinfo {year} {2019})}\BibitemShut {NoStop}%
\bibitem [{\citenamefont {Pozas-Kerstjens}(2019)}]{pozas2019quantum}%
  \BibitemOpen
  \bibfield  {author} {\bibinfo {author} {\bibfnamefont {A.}~\bibnamefont
  {Pozas-Kerstjens}},\ }\emph {\bibinfo {title} {Quantum information outside
  quantum information}},\ \href
  {https://www.tdx.cat/handle/10803/667696#page=1} {Ph.D. thesis},\ \bibinfo
  {school} {Universitat Polit{\`e}cnica de Catalunya} (\bibinfo {year}
  {2019})\BibitemShut {NoStop}%
\bibitem [{\citenamefont {Pironio}\ \emph {et~al.}(2010)\citenamefont
  {Pironio}, \citenamefont {Navascu{\'e}s},\ and\ \citenamefont
  {Acin}}]{pironio2010}%
  \BibitemOpen
  \bibfield  {author} {\bibinfo {author} {\bibfnamefont {S.}~\bibnamefont
  {Pironio}}, \bibinfo {author} {\bibfnamefont {M.}~\bibnamefont
  {Navascu{\'e}s}}, \ and\ \bibinfo {author} {\bibfnamefont {A.}~\bibnamefont
  {Acin}},\ }\href {\doibase https://doi.org/10.1137/090760155} {\bibfield
  {journal} {\bibinfo  {journal} {SIAM J. Optim.}\ }\textbf {\bibinfo {volume}
  {20}},\ \bibinfo {pages} {2157} (\bibinfo {year} {2010})}\BibitemShut
  {NoStop}%
\bibitem [{Note1()}]{Note1}%
  \BibitemOpen
  \bibinfo {note} {While the scalar extension hierarchy was formulated earlier,
  Ref.~\cite {klep2023state} shows also completeness of the hierarchy. We here
  use the term state polynomial optimization, as it highlights the connections
  to polynomial optimization best. In any case, the two terms can be used
  interchangeably.}\BibitemShut {Stop}%
\bibitem [{\citenamefont {Galli}\ and\ \citenamefont
  {Letchford}(2017)}]{GALLI2017159}%
  \BibitemOpen
  \bibfield  {author} {\bibinfo {author} {\bibfnamefont {L.}~\bibnamefont
  {Galli}}\ and\ \bibinfo {author} {\bibfnamefont {A.~N.}\ \bibnamefont
  {Letchford}},\ }\href {\doibase https://doi.org/10.1016/j.disopt.2017.04.001}
  {\bibfield  {journal} {\bibinfo  {journal} {Discret. Optim.}\ }\textbf
  {\bibinfo {volume} {25}},\ \bibinfo {pages} {159} (\bibinfo {year}
  {2017})}\BibitemShut {NoStop}%
\bibitem [{Note2()}]{Note2}%
  \BibitemOpen
  \bibinfo {note} {Interestingly, we observe a similar numerical behaviour with
  the sequence $\{\langle (A_i A_j)^\protect \dag \rangle _\varrho A_i A_j\}_{0
  \leq i < j \leq n}$.}\BibitemShut {Stop}%
\bibitem [{\citenamefont {Tavakoli}\ \emph {et~al.}(2022)\citenamefont
  {Tavakoli}, \citenamefont {Pozas-Kerstjens}, \citenamefont {Luo},\ and\
  \citenamefont {Renou}}]{Tavakoli_2022}%
  \BibitemOpen
  \bibfield  {author} {\bibinfo {author} {\bibfnamefont {A.}~\bibnamefont
  {Tavakoli}}, \bibinfo {author} {\bibfnamefont {A.}~\bibnamefont
  {Pozas-Kerstjens}}, \bibinfo {author} {\bibfnamefont {M.-X.}\ \bibnamefont
  {Luo}}, \ and\ \bibinfo {author} {\bibfnamefont {M.-O.}\ \bibnamefont
  {Renou}},\ }\href {\doibase 10.1088/1361-6633/ac41bb} {\bibfield  {journal}
  {\bibinfo  {journal} {Rep. Prog. Phys.}\ }\textbf {\bibinfo {volume} {85}},\
  \bibinfo {pages} {056001} (\bibinfo {year} {2022})}\BibitemShut {NoStop}%
\bibitem [{\citenamefont {Maassen}\ and\ \citenamefont
  {Uffink}(1988)}]{PhysRevLett.60.1103}%
  \BibitemOpen
  \bibfield  {author} {\bibinfo {author} {\bibfnamefont {H.}~\bibnamefont
  {Maassen}}\ and\ \bibinfo {author} {\bibfnamefont {J.~B.~M.}\ \bibnamefont
  {Uffink}},\ }\href {\doibase 10.1103/PhysRevLett.60.1103} {\bibfield
  {journal} {\bibinfo  {journal} {Phys. Rev. Lett.}\ }\textbf {\bibinfo
  {volume} {60}},\ \bibinfo {pages} {1103} (\bibinfo {year}
  {1988})}\BibitemShut {NoStop}%
\bibitem [{\citenamefont {Wehner}\ and\ \citenamefont
  {Winter}(2008)}]{Wehner_2008}%
  \BibitemOpen
  \bibfield  {author} {\bibinfo {author} {\bibfnamefont {S.}~\bibnamefont
  {Wehner}}\ and\ \bibinfo {author} {\bibfnamefont {A.}~\bibnamefont
  {Winter}},\ }\href {\doibase 10.1063/1.2943685} {\bibfield  {journal}
  {\bibinfo  {journal} {J. Math. Phys.}\ }\textbf {\bibinfo {volume} {49}},\
  \bibinfo {pages} {062105} (\bibinfo {year} {2008})}\BibitemShut {NoStop}%
\bibitem [{\citenamefont {Pucha{\l}a}\ \emph {et~al.}(2013)\citenamefont
  {Pucha{\l}a}, \citenamefont {Rudnicki},\ and\ \citenamefont
  {{\.Z}yczkowski}}]{puchala2013majorization}%
  \BibitemOpen
  \bibfield  {author} {\bibinfo {author} {\bibfnamefont {Z.}~\bibnamefont
  {Pucha{\l}a}}, \bibinfo {author} {\bibfnamefont {{\L}.}~\bibnamefont
  {Rudnicki}}, \ and\ \bibinfo {author} {\bibfnamefont {K.}~\bibnamefont
  {{\.Z}yczkowski}},\ }\href {\doibase 10.1088/1751-8113/46/27/272002}
  {\bibfield  {journal} {\bibinfo  {journal} {J. Phys. A Math. Theor.}\
  }\textbf {\bibinfo {volume} {46}},\ \bibinfo {pages} {272002} (\bibinfo
  {year} {2013})}\BibitemShut {NoStop}%
\bibitem [{\citenamefont {Korzekwa}\ \emph {et~al.}(2014)\citenamefont
  {Korzekwa}, \citenamefont {Lostaglio}, \citenamefont {Jennings},\ and\
  \citenamefont {Rudolph}}]{korzekwa2014quantum}%
  \BibitemOpen
  \bibfield  {author} {\bibinfo {author} {\bibfnamefont {K.}~\bibnamefont
  {Korzekwa}}, \bibinfo {author} {\bibfnamefont {M.}~\bibnamefont {Lostaglio}},
  \bibinfo {author} {\bibfnamefont {D.}~\bibnamefont {Jennings}}, \ and\
  \bibinfo {author} {\bibfnamefont {T.}~\bibnamefont {Rudolph}},\ }\href
  {\doibase https://doi.org/10.1103/PhysRevA.89.042122} {\bibfield  {journal}
  {\bibinfo  {journal} {Phys. Rev. A}\ }\textbf {\bibinfo {volume} {89}},\
  \bibinfo {pages} {042122} (\bibinfo {year} {2014})}\BibitemShut {NoStop}%
\bibitem [{\citenamefont {Zhao}\ \emph {et~al.}(2019)\citenamefont {Zhao},
  \citenamefont {Xiang}, \citenamefont {Hu}, \citenamefont {Liu}, \citenamefont
  {Li}, \citenamefont {Guo}, \citenamefont {Schwonnek},\ and\ \citenamefont
  {Wolf}}]{Zhao_2019}%
  \BibitemOpen
  \bibfield  {author} {\bibinfo {author} {\bibfnamefont {Y.-Y.}\ \bibnamefont
  {Zhao}}, \bibinfo {author} {\bibfnamefont {G.-Y.}\ \bibnamefont {Xiang}},
  \bibinfo {author} {\bibfnamefont {X.-M.}\ \bibnamefont {Hu}}, \bibinfo
  {author} {\bibfnamefont {B.-H.}\ \bibnamefont {Liu}}, \bibinfo {author}
  {\bibfnamefont {C.-F.}\ \bibnamefont {Li}}, \bibinfo {author} {\bibfnamefont
  {G.-C.}\ \bibnamefont {Guo}}, \bibinfo {author} {\bibfnamefont
  {R.}~\bibnamefont {Schwonnek}}, \ and\ \bibinfo {author} {\bibfnamefont
  {R.}~\bibnamefont {Wolf}},\ }\href {\doibase 10.1103/PhysRevLett.122.220401}
  {\bibfield  {journal} {\bibinfo  {journal} {Phys. Rev. Lett.}\ }\textbf
  {\bibinfo {volume} {122}},\ \bibinfo {pages} {220401} (\bibinfo {year}
  {2019})}\BibitemShut {NoStop}%
\bibitem [{\citenamefont {Rotundo}\ and\ \citenamefont
  {Schwonnek}(2024)}]{rotundo2023entropic}%
  \BibitemOpen
  \bibfield  {author} {\bibinfo {author} {\bibfnamefont {A.~F.}\ \bibnamefont
  {Rotundo}}\ and\ \bibinfo {author} {\bibfnamefont {R.}~\bibnamefont
  {Schwonnek}},\ }\href {\doibase 10.1103/PhysRevResearch.6.033043} {\bibfield
  {journal} {\bibinfo  {journal} {Phys. Rev. Res.}\ }\textbf {\bibinfo {volume}
  {6}},\ \bibinfo {pages} {033043} (\bibinfo {year} {2024})}\BibitemShut
  {NoStop}%
\bibitem [{\citenamefont {Wang}\ and\ \citenamefont
  {Magron}(2021)}]{wang2021exploiting}%
  \BibitemOpen
  \bibfield  {author} {\bibinfo {author} {\bibfnamefont {J.}~\bibnamefont
  {Wang}}\ and\ \bibinfo {author} {\bibfnamefont {V.}~\bibnamefont {Magron}},\
  }\href {\doibase https://doi.org/10.1007/s10589-021-00301-7} {\bibfield
  {journal} {\bibinfo  {journal} {Comput. Optim. Appl.}\ }\textbf {\bibinfo
  {volume} {80}},\ \bibinfo {pages} {483} (\bibinfo {year} {2021})}\BibitemShut
  {NoStop}%
\bibitem [{\citenamefont {Ioannou}\ and\ \citenamefont
  {Rosset}(2021)}]{ioannou2021noncommutative}%
  \BibitemOpen
  \bibfield  {author} {\bibinfo {author} {\bibfnamefont {M.}~\bibnamefont
  {Ioannou}}\ and\ \bibinfo {author} {\bibfnamefont {D.}~\bibnamefont
  {Rosset}},\ }\href@noop {} {\enquote {\bibinfo {title} {Noncommutative
  polynomial optimization under symmetry},}\ } (\bibinfo {year} {2021}),\
  \Eprint {http://arxiv.org/abs/2112.10803} {arXiv:2112.10803} \BibitemShut
  {NoStop}%
\bibitem [{\citenamefont {Bachoc}\ \emph {et~al.}(2012)\citenamefont {Bachoc},
  \citenamefont {Gijswijt}, \citenamefont {Schrijver},\ and\ \citenamefont
  {Vallentin}}]{bachoc2012invariant}%
  \BibitemOpen
  \bibfield  {author} {\bibinfo {author} {\bibfnamefont {C.}~\bibnamefont
  {Bachoc}}, \bibinfo {author} {\bibfnamefont {D.~C.}\ \bibnamefont
  {Gijswijt}}, \bibinfo {author} {\bibfnamefont {A.}~\bibnamefont {Schrijver}},
  \ and\ \bibinfo {author} {\bibfnamefont {F.}~\bibnamefont {Vallentin}},\
  }\enquote {\bibinfo {title} {Invariant semidefinite programs},}\ in\ \href
  {https://link.springer.com/book/10.1007/978-1-4614-0769-0} {\emph {\bibinfo
  {booktitle} {{Handbook on semidefinite, conic and polynomial
  optimization}}}}\ (\bibinfo  {publisher} {Springer},\ \bibinfo {year}
  {2012})\ pp.\ \bibinfo {pages} {219--269}\BibitemShut {NoStop}%
\bibitem [{\citenamefont {Permenter}\ and\ \citenamefont
  {Parrilo}(2020)}]{permenter2020dimension}%
  \BibitemOpen
  \bibfield  {author} {\bibinfo {author} {\bibfnamefont {F.}~\bibnamefont
  {Permenter}}\ and\ \bibinfo {author} {\bibfnamefont {P.~A.}\ \bibnamefont
  {Parrilo}},\ }\href {\doibase https://doi.org/10.1007/s10107-019-01372-5}
  {\bibfield  {journal} {\bibinfo  {journal} {Math. Program.}\ }\textbf
  {\bibinfo {volume} {181}},\ \bibinfo {pages} {51} (\bibinfo {year}
  {2020})}\BibitemShut {NoStop}%
\bibitem [{\citenamefont {Brosch}\ and\ \citenamefont
  {de~Klerk}(2022)}]{brosch2022jordan}%
  \BibitemOpen
  \bibfield  {author} {\bibinfo {author} {\bibfnamefont {D.}~\bibnamefont
  {Brosch}}\ and\ \bibinfo {author} {\bibfnamefont {E.}~\bibnamefont
  {de~Klerk}},\ }\href {\doibase https://doi.org/10.1080/10556788.2021.2022146}
  {\bibfield  {journal} {\bibinfo  {journal} {Optim. Methods Softw.}\ }\textbf
  {\bibinfo {volume} {37}},\ \bibinfo {pages} {2001} (\bibinfo {year}
  {2022})}\BibitemShut {NoStop}%
\bibitem [{\citenamefont {Ligthart}\ and\ \citenamefont
  {Gross}(2023)}]{ligthart2022inflation}%
  \BibitemOpen
  \bibfield  {author} {\bibinfo {author} {\bibfnamefont {L.~T.}\ \bibnamefont
  {Ligthart}}\ and\ \bibinfo {author} {\bibfnamefont {D.}~\bibnamefont
  {Gross}},\ }\href {\doibase 10.1063/5.0143792} {\bibfield  {journal}
  {\bibinfo  {journal} {J. Math. Phys.}\ }\textbf {\bibinfo {volume} {64}},\
  \bibinfo {pages} {072201} (\bibinfo {year} {2023})}\BibitemShut {NoStop}%
\bibitem [{\citenamefont {Gastineau-Hills}(1982)}]{gastineau1982quasi}%
  \BibitemOpen
  \bibfield  {author} {\bibinfo {author} {\bibfnamefont {H.}~\bibnamefont
  {Gastineau-Hills}},\ }\href {\doibase 10.1017/S1446788700024368} {\bibfield
  {journal} {\bibinfo  {journal} {J. Aust. Math. Soc.}\ }\textbf {\bibinfo
  {volume} {32}},\ \bibinfo {pages} {1} (\bibinfo {year} {1982})}\BibitemShut
  {NoStop}%
\bibitem [{\citenamefont {Panov}(2001)}]{panov1998quantum}%
  \BibitemOpen
  \bibfield  {author} {\bibinfo {author} {\bibfnamefont {A.~N.}\ \bibnamefont
  {Panov}},\ }\href {\doibase https://doi.org/10.1023/A:1010264331776}
  {\bibfield  {journal} {\bibinfo  {journal} {Mathematical Notes}\ }\textbf
  {\bibinfo {volume} {69}},\ \bibinfo {pages} {537} (\bibinfo {year}
  {2001})}\BibitemShut {NoStop}%
\bibitem [{\citenamefont {Vourdas}(2004)}]{Vourdas_2004}%
  \BibitemOpen
  \bibfield  {author} {\bibinfo {author} {\bibfnamefont {A.}~\bibnamefont
  {Vourdas}},\ }\href {\doibase 10.1088/0034-4885/67/3/R03} {\bibfield
  {journal} {\bibinfo  {journal} {Rep. Prog. Phys.}\ }\textbf {\bibinfo
  {volume} {67}},\ \bibinfo {pages} {267} (\bibinfo {year} {2004})}\BibitemShut
  {NoStop}%
\bibitem [{\citenamefont {Klappenecker}\ and\ \citenamefont
  {R{\"{o}}tteler}(2002)}]{klappenecker2001stabilizer}%
  \BibitemOpen
  \bibfield  {author} {\bibinfo {author} {\bibfnamefont {A.}~\bibnamefont
  {Klappenecker}}\ and\ \bibinfo {author} {\bibfnamefont {M.}~\bibnamefont
  {R{\"{o}}tteler}},\ }\href {\doibase 10.1109/TIT.2002.800471} {\bibfield
  {journal} {\bibinfo  {journal} {{IEEE} Trans. Inf. Theory}\ }\textbf
  {\bibinfo {volume} {48}},\ \bibinfo {pages} {2392} (\bibinfo {year}
  {2002})}\BibitemShut {NoStop}%
\bibitem [{\citenamefont {\.Zukowski}\ and\ \citenamefont
  {Brukner}(2002)}]{PhysRevLett.88.210401}%
  \BibitemOpen
  \bibfield  {author} {\bibinfo {author} {\bibfnamefont {M.}~\bibnamefont
  {\.Zukowski}}\ and\ \bibinfo {author} {\bibfnamefont {{\v C}.}~\bibnamefont
  {Brukner}},\ }\href {\doibase 10.1103/PhysRevLett.88.210401} {\bibfield
  {journal} {\bibinfo  {journal} {Phys. Rev. Lett.}\ }\textbf {\bibinfo
  {volume} {88}},\ \bibinfo {pages} {210401} (\bibinfo {year}
  {2002})}\BibitemShut {NoStop}%
\bibitem [{\citenamefont {Karp}(1972)}]{Karp1972}%
  \BibitemOpen
  \bibfield  {author} {\bibinfo {author} {\bibfnamefont {R.~M.}\ \bibnamefont
  {Karp}},\ }\enquote {\bibinfo {title} {Reducibility among combinatorial
  problems},}\ in\ \href {\doibase 10.1007/978-1-4684-2001-2_9} {\emph
  {\bibinfo {booktitle} {Complexity of Computer Computations}}},\ \bibinfo
  {series and number} {IBM Research Symposia Series},\ \bibinfo {editor}
  {edited by\ \bibinfo {editor} {\bibfnamefont {R.~E.}\ \bibnamefont {Miller}},
  \bibinfo {editor} {\bibfnamefont {J.~W.}\ \bibnamefont {Thatcher}}, \ and\
  \bibinfo {editor} {\bibfnamefont {J.~D.}\ \bibnamefont {Bohlinger}}}\
  (\bibinfo  {publisher} {Springer US},\ \bibinfo {address} {Boston, MA},\
  \bibinfo {year} {1972})\ pp.\ \bibinfo {pages} {85--103}\BibitemShut
  {NoStop}%
\bibitem [{\citenamefont {Knuth}(1994)}]{knuth1993sandwich}%
  \BibitemOpen
  \bibfield  {author} {\bibinfo {author} {\bibfnamefont {D.~E.}\ \bibnamefont
  {Knuth}},\ }\href {http://eudml.org/doc/118559} {\bibfield  {journal}
  {\bibinfo  {journal} {The Electronic Journal of Combinatorics}\ }\textbf
  {\bibinfo {volume} {1}} (\bibinfo {year} {1994})}\BibitemShut {NoStop}%
\bibitem [{\citenamefont {Lov{\'a}sz}(1979)}]{lovasz1979}%
  \BibitemOpen
  \bibfield  {author} {\bibinfo {author} {\bibfnamefont {L.}~\bibnamefont
  {Lov{\'a}sz}},\ }\href {\doibase 10.1109/TIT.1979.1055985} {\bibfield
  {journal} {\bibinfo  {journal} {IEEE Trans. Inf. Theory}\ }\textbf {\bibinfo
  {volume} {25}},\ \bibinfo {pages} {1} (\bibinfo {year} {1979})}\BibitemShut
  {NoStop}%
\bibitem [{\citenamefont {Christandl}\ \emph {et~al.}(2007)\citenamefont
  {Christandl}, \citenamefont {K{\"o}nig}, \citenamefont {Mitchison},\ and\
  \citenamefont {Renner}}]{Christandl2007}%
  \BibitemOpen
  \bibfield  {author} {\bibinfo {author} {\bibfnamefont {M.}~\bibnamefont
  {Christandl}}, \bibinfo {author} {\bibfnamefont {R.}~\bibnamefont
  {K{\"o}nig}}, \bibinfo {author} {\bibfnamefont {G.}~\bibnamefont
  {Mitchison}}, \ and\ \bibinfo {author} {\bibfnamefont {R.}~\bibnamefont
  {Renner}},\ }\href {\doibase 10.1007/s00220-007-0189-3} {\bibfield  {journal}
  {\bibinfo  {journal} {Commun. Math. Phys.}\ }\textbf {\bibinfo {volume}
  {273}},\ \bibinfo {pages} {473} (\bibinfo {year} {2007})}\BibitemShut
  {NoStop}%
\bibitem [{\citenamefont {Ketkar}\ \emph {et~al.}(2006)\citenamefont {Ketkar},
  \citenamefont {Klappenecker}, \citenamefont {Kumar},\ and\ \citenamefont
  {Sarvepalli}}]{ketkar2006nonbinary}%
  \BibitemOpen
  \bibfield  {author} {\bibinfo {author} {\bibfnamefont {A.}~\bibnamefont
  {Ketkar}}, \bibinfo {author} {\bibfnamefont {A.}~\bibnamefont
  {Klappenecker}}, \bibinfo {author} {\bibfnamefont {S.}~\bibnamefont {Kumar}},
  \ and\ \bibinfo {author} {\bibfnamefont {P.~K.}\ \bibnamefont {Sarvepalli}},\
  }\href {\doibase 10.1109/TIT.2006.883612} {\bibfield  {journal} {\bibinfo
  {journal} {IEEE Trans. Inf. Theor.}\ }\textbf {\bibinfo {volume} {52}},\
  \bibinfo {pages} {4892?4914} (\bibinfo {year} {2006})}\BibitemShut {NoStop}%
\bibitem [{\citenamefont {Scott}(2004)}]{scott2004multipartite}%
  \BibitemOpen
  \bibfield  {author} {\bibinfo {author} {\bibfnamefont {A.~J.}\ \bibnamefont
  {Scott}},\ }\href {\doibase https://doi.org/10.1103/PhysRevA.69.052330}
  {\bibfield  {journal} {\bibinfo  {journal} {Phys. Rev. A}\ }\textbf {\bibinfo
  {volume} {69}},\ \bibinfo {pages} {052330} (\bibinfo {year}
  {2004})}\BibitemShut {NoStop}%
\end{thebibliography}
\end{document}